\documentclass[prl,twocolumn,amsmath,amssymb,floatfix,showpacs]{revtex4}

\usepackage{graphicx}

\usepackage{bm}

\usepackage{amssymb}

\usepackage{color}
\usepackage{epsfig}

\begin{document}

\title{Decays in Quantum Hierarchical Models}

\author{Ariel Amir, Yuval Oreg, Yoseph Imry}

\affiliation { Department of Condensed Matter Physics, Weizmann
Institute of Science, Rehovot, 76100, Israel\\}

\newcommand{\be}{\begin{equation}}
\newcommand{\ee}{\end{equation}}


\begin{abstract}

 We study the dynamics of a simple model for quantum decay, where a single state is coupled to a set of discrete states,
 the pseudo continuum,
 each coupled to a real continuum of states. We find that for constant matrix elements
 between the single state and the pseudo continuum the decay occurs via one state in a certain region of the
 parameters, involving the Dicke and quantum Zeno effects. When the matrix elements are random several cases are identified.
 For a pseudo continuum with small
 bandwidth there are weakly damped oscillations in the probability to be in the initial single
 state.
  For intermediate bandwidth one finds mesoscopic
fluctuations in the probability with amplitude inversely
proportional to the square root of the volume of the pseudo
continuum space. They last for a long time compared to the
non-random case.
\end{abstract}
\maketitle
 The problem of the decay of an excitation into a
continuum is a fundamental problem in quantum mechanics~
\cite{cohen_tanoudji_decay}, appearing in numerous fields of
physics. A natural hierarchy of couplings occurs in many physical
systems,

For example, a spin of a nucleus may be coupled to the
electromagnetic modes of a cavity in which it is situated, and these
in turn may be coupled to the modes of a larger box or the vacuum
\cite {purcell}. Hierarchical systems were studied  \cite
{{exp_microwave},
{exp_nuclear},{TLS1},{gurvitz2},{zelevinksy_RMT},{zelevinksy_RMT2},{silvestrov},{brouwer1}}
but the time dependence for the model we present was not
investigated.

\begin{figure}[htbp]
\centerline{\includegraphics[height=2in]{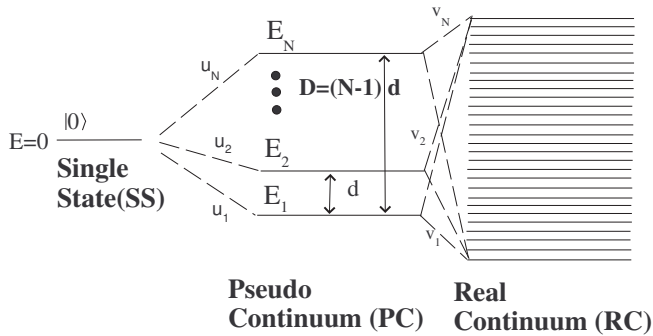}} \caption {The
model consists of a single state (SS) coupled to a discrete number
of states which form the pseudo continuum (PC), each of which is
coupled to a dense and broad 'real' continuum (RC).}
\label{model_fig}
\end{figure}

 \emph{Definition of
the problem and main results}.- Our model is as follows: The single
state (SS) $|0\rangle$ is coupled with matrix elements $u_\mu$ to
$N$ states which form the pseudo continuum (PC), of bandwidth $D$
and constant level spacing $d=\frac{D}{N-1}$. Each of these is
coupled with a matrix element ${v_\mu}$ to a broad and dense real
continuum (RC), see Fig. \ref {model_fig}. Notice that for each
state of the PC the matrix element is identical for all states in
the continuum, see footnote~\cite
{remark_about_single_channel_leads}. For convenience we take
$E_0=0$, and all matrix elements to be real, assuming time reversal
symmetry. The probability amplitude of finding the system in state
$|0\rangle$ is denoted by $A_0(t)$. Starting with $A_0(0)=1$, we
evaluate $A_0(t)$ at $t>0$. The Hamiltonian of the system is:

\be H_{\rm{total}}= \sum_{\mu=1}^N E_\mu |\mu \rangle \langle \mu|+
\sum_{k=1}^M \tilde{E_k} |k \rangle \langle k|+ H_{\rm{couplings}}
\label {H_} ,\ee with $N \gg 1$, and $M \rightarrow \infty $
(keeping the level spacing infinitesimal and the bandwidth large)
and

\be H_{\rm{couplings}}= \sum_{\mu=1}^N u_\mu |0 \rangle \langle \mu|
+ \sum_{\mu,k} v_\mu |\mu \rangle \langle k|+ h.c. \space \space.\ee

We will first show that in the case of constant couplings $u_\mu=u$,
$v_\mu=v$ the system decay is dominantly exponential with rate: \be
\Gamma= \frac{u^2}{\gamma + \frac{d}{\pi}} , \label {generalized_eq}
\ee where $\gamma= \pi {v^2}\nu_c$, $\nu_c$ being the RC density of
states \cite{remark_about_alternating_sign_case}. As the coupling to
the RC becomes stronger, the rate decreases, a phenomenon referred
to as the quantum Zeno effect \cite { {zeno1},{zeno2}}. A
diagonalization of Hamlitonian (\ref {H_}) for the case $\gamma \gg
d$ yields a very wide state, similar to the Dicke state \cite
{dicke}, and another wide state, which is the most relevant for the
decay of the system. The other $N-1$ states acquire a small width
and are not pertinent for the dynamics of the system until long
times of order $\frac{\gamma}{u^2}\log(\frac{\gamma}{d})$.

\begin{figure}[h]
\begin{center}
$\begin{array}{c@{\hspace{0.00in}}c} \multicolumn{1}{l}{\mbox{\bf
(a)}} &
    \multicolumn{1}{l}{\mbox{\bf (b) }} \\ [-0.0cm]
\epsfxsize=1.6in \epsffile{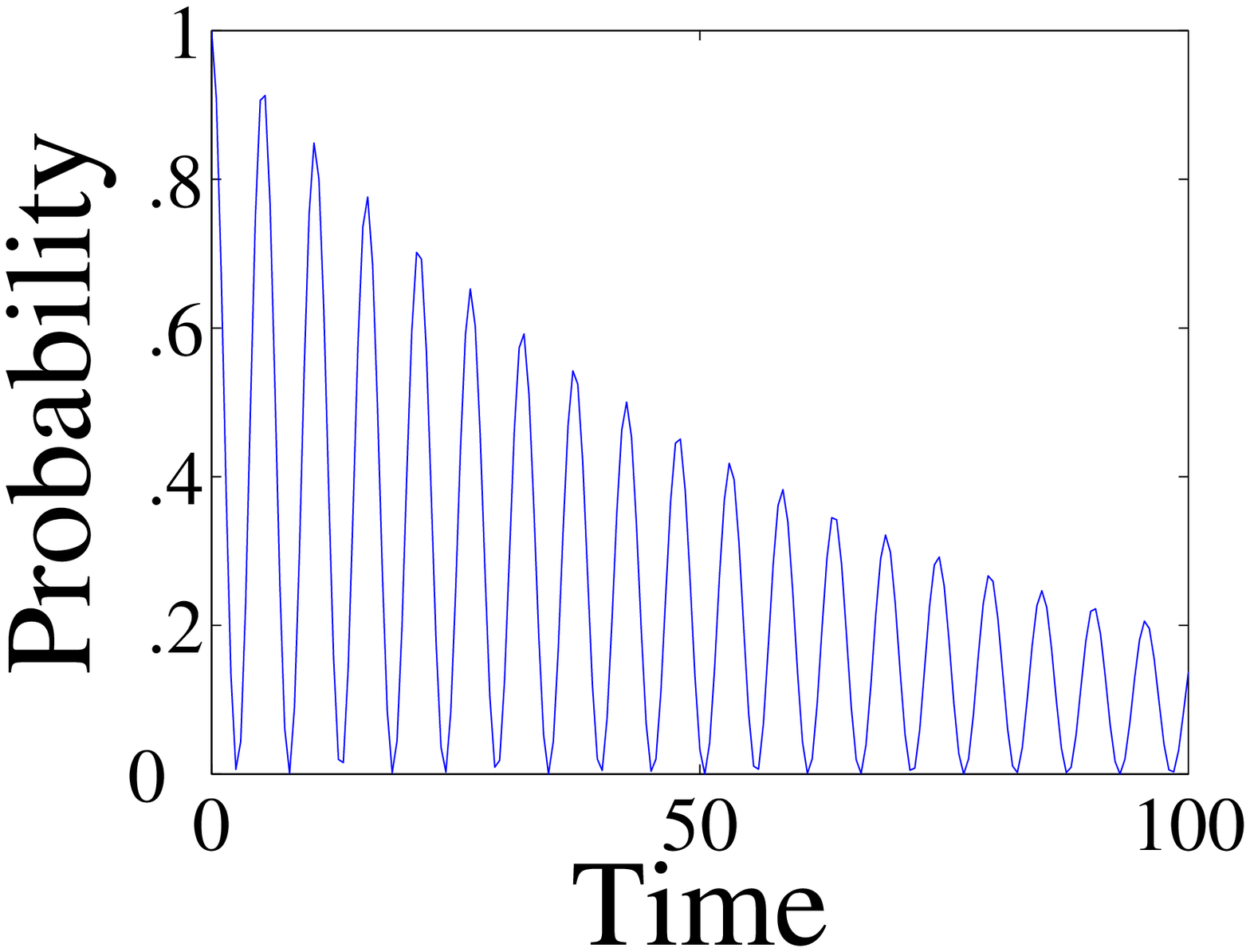} &
    \epsfxsize=1.6in
    \epsffile{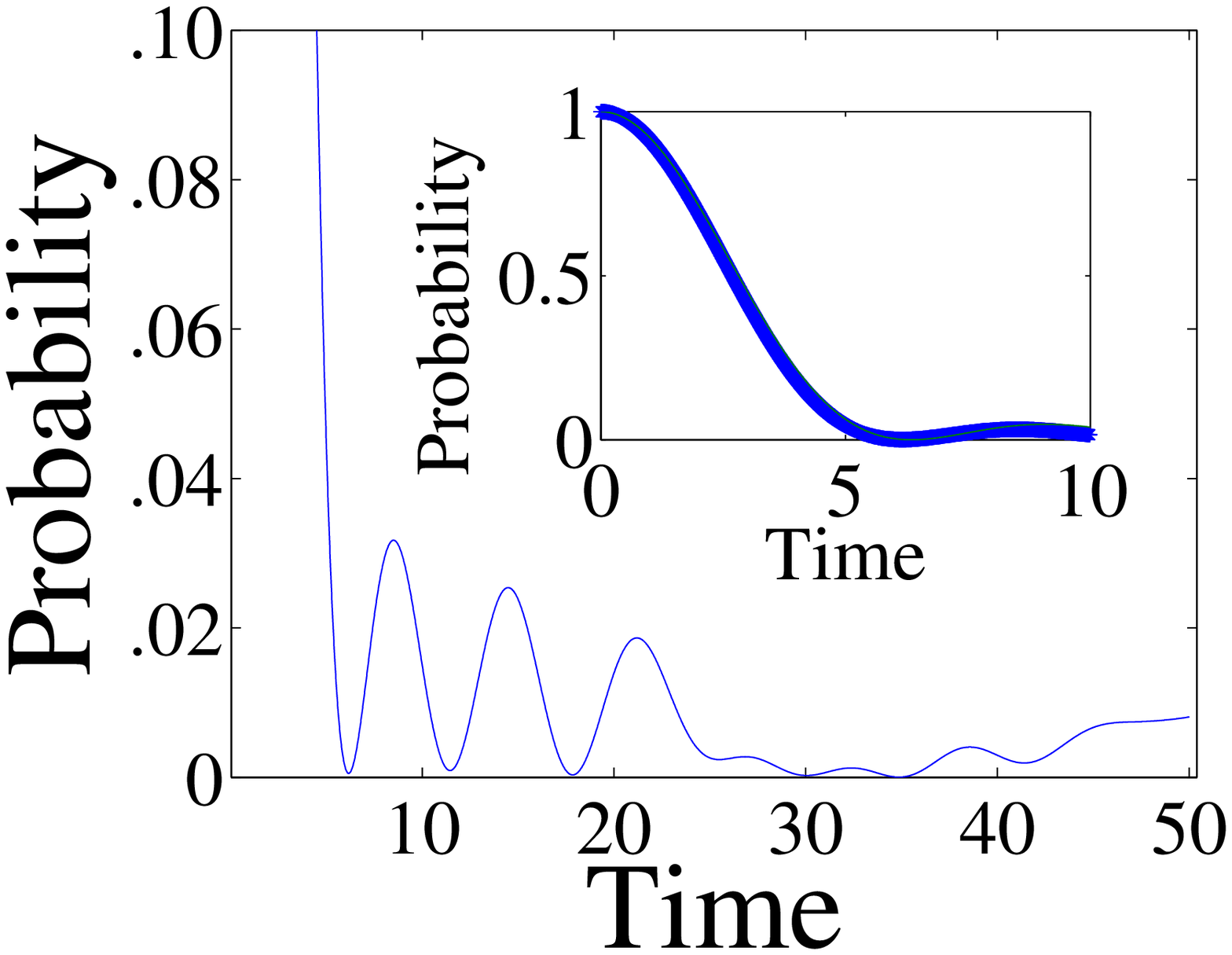} \\ [0.2 cm]
\mbox{\bf Zero level spacing  } & \mbox{\bf Constant level spacing }
\end{array}$
\end{center}
\caption{ Decay of a SS coupled to a PC with random matrix elements
$u_\mu$ (see Eq. (\ref{H})), uniformly distributed with standard
deviation 0.1. For $D=0$ (left) $\gamma=0.1$, $N=21$, while for the
non degenerate case $D=1$ (right) $\gamma=1$, $N=101$. The
degenerate case shows damped oscillations, while the finite
bandwidth case shows a sharp initial decay followed by mesoscopic
fluctuations. The inset shows the short-time
fit to a sinc function. 
}
\label{random3}
\end{figure}

When the couplings to the RC are random we identify four regimes.
The boundaries between them are controlled by the typical matrix
element $\bar{u} \sim \frac{u_0}{\sqrt{N}}$. For $u_0<d$ there are
decaying Rabi oscillations via one state. For $u_0<D$ Fermi's golden
rule (FGR) is obtained, while for $u_0 \sim D$ novel mesoscopic
fluctuations with amplitude $\sim \frac{1}{\sqrt{N}}$ appear, see
Fig. \ref {random3}. For the almost degenerate case, i.e., $u_0>D$,
we find decaying oscillations via a linear combination of many
states in the PC.

\emph{Physical realizations}.- The model (\ref {H_}) applies, for
example, to a small quantum dot coupled to one or more quantum dots,
each of which is coupled to a lead. One can study the time
dependence of an injected electron's probability to remain in the
dot, assuming that all relevant levels in the dots are empty. In
some cases the matrix elements $u_\mu$ connecting two dots can be
taken as constant \cite {quantum_dot_physical_realization}. For the
more generic case of a single dot coupled to a disordered or
sufficiently distorted larger quantum dot, $u_\mu$ do not have the
same sign, and we take them to be random. These solid state
implementations have close analogies when one replaces the quantum
dots by atoms in optical cavities \cite{purcell}.

\emph{Derivation of the results}.- Eliminating the amplitudes in the
RC after Laplace transforming the equations of motion, the dynamics
of the other $N+1$ amplitudes in the SS and the PC are formulated in
terms of a $N+1$ by $N+1$ non-hermitian matrix, which is the matrix
describing the original matrix elements between these states, plus a
matrix element $- i\gamma_{\mu \nu}=-i \pi \nu_c {v_\mu} {v_\nu} $
\cite { {remark_about_single_channel_leads},{zelevinksy_unstable}}.
The reduced Hamiltonian for the system is:

\be H= \sum_{\mu=1}^N E_\mu |\mu \rangle \langle \mu|+ u_\mu (|0
\rangle \langle \mu| + |\mu \rangle \langle 0|)-\sum_{\mu,\nu} i
\gamma_{\mu \nu} |\mu \rangle \langle \nu| . \label {H}\ee  From now
on we shall assume $v_\mu$ to be constant, and therefore
$\gamma_{\mu \nu}\equiv\gamma$.

This leads to the (exact) eigenvalue equation:

\be \sum \frac{u_\mu ^2}{\lambda - E_\mu} - i \gamma \Sigma_1 \frac
{u_\mu}{\lambda - E_\mu} = \lambda , \label {eigenval_eq} \ee with
$\Sigma_1= \frac{\sum \frac{u_\mu}{\lambda-E_\mu} }{i \gamma \sum
\frac{1}{\lambda - E_\mu}+1}$.

The generic form for the eigenvectors is:

\be |V_n\rangle = |0 \rangle + \sum_\mu \frac{u_\mu - i \gamma
\Sigma_1}{\lambda_n- E_\mu} |\mu \rangle \label {eigenvectors} ,\ee
where $\lambda_n$ is the corresponding eigenvalue.

Although $H$ is non-hermitian, we can still decompose the initial
state as a superposition of its eigenvectors \cite
{pertubation_non_hermitian2}: \be A_0(t)= \sum_n C_n e^{-i \lambda_n
t} \langle 0|V_n\rangle \label {time_evolution}. \ee Notice that
even when we normalize the states $|V_n \rangle$, the coefficients
$C_n$ are not the usual projections ${\langle V_n|0\rangle}$
\cite{{pertubation_non_hermitian2}, {remark_about_non_hermitian}}.

We shall now use this formalism to study the cases of constant and
random matrix elements.

 \emph{A. Constant matrix elements}.- First we take
 the matrix element between the initial SS and the levels of the PC to be a constant $u$ \cite {quantum_dot_physical_realization}.
%

To analyze the decay, let us write the equations of motion for the
amplitudes following from Eq. (\ref {H}):

\be i \frac{d A_\mu}{dt} =  u A_0 -i \gamma \sum_\nu A_\nu +E_\mu
A_\mu .\label {time_evolve}\ee

where $A_\mu$ is the amplitude of state $\mu$.

Upon Laplace transforming Eq. (\ref {time_evolve}), we obtain:

\begin{eqnarray}
i \omega A_0 = i + u \sum_\mu A_\mu, \\
i \omega A_\mu =  u A_0 -i \gamma \sum_\nu A_\nu +E_\mu A_\mu.
\end{eqnarray}

For $D \gg \omega \gg d$ and for the SS with energy far enough from
the edges of the PC band, we can approximate the sum $\sum
\frac{1}{i \omega -E_\mu}$ by $-\frac{i \pi}{d}$, which leads to the
result:

\be A_0= \frac{1}{\omega + \Gamma},\ee with $\Gamma$ given by Eq.
(\ref {generalized_eq}). The inverse Laplace transform gives the
exponential decay, in a large time window, which for $\gamma \ll d$
is given by $\frac{1}{D}<t<\frac{1}{d}$.

We shall now analyze the structure of the eigenstates yielding the
result of Eq. (\ref {generalized_eq}).

For the limit $\gamma=0$, we have the discrete Wigner-Weisskopf
problem \cite {cohen_tanoudji_decay},
 and Eq. (\ref {generalized_eq}) reduces to
FGR. In that case the eigenvalues are real (since the Hamiltonian is
hermitian), and it is the superposition of many eigenvectors that
gives rise to the decay (for intermediate times).

\begin{figure}[htbp]
\centerline{\includegraphics[height=2.56in]{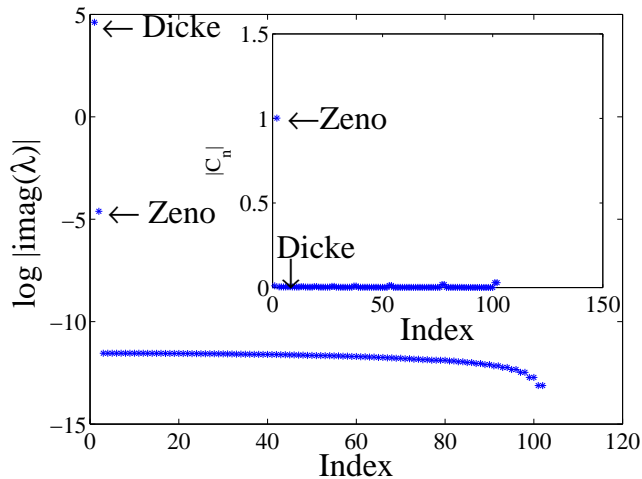}} 
\caption {Imaginary part of the eigenvalues of Hamiltonian
(\ref{H}), with constant matrix elements. u is 0.1, D is 1, $\gamma$
is 1, and there are 101 states in the PC. The inset shows $|C_n|$
define in Eq. (\ref {time_evolution}). The weight is concentrated on
one eigenvector, the Zeno state, decaying faster than all the rest
of the eigenvalues except for the Dicke state. } \label{eigenvals}
\end{figure}

As $\gamma$ increases and reaches the regime $\gamma \gg d$, the
behavior is changed and one state  completely dominates the decay of
the system. In this regime there is a fast decaying eigenvector
approximately of the form $|0 \rangle + x \sum_j |j \rangle $, with
eigenvalue $\lambda \approx -i \gamma N$ and $ x \approx
\frac{\lambda}{u N} $. This is related to the Dicke effect, where a
coherent sum of many states with equal amplitudes is also present
\cite {dicke}. We also find an additional eigenvalue $-i \frac{u^2}{
\gamma}$. Since an increase in $\gamma$ causes a smaller decay rate,
we are motivated to call the corresponding eigenvector the Zeno
state. Using perturbation theory \cite{pertubation_non_hermitian2}
one may show that in this case the coefficients $C_n$ of Eq. (\ref
{time_evolution}) are nearly unity for the Zeno state, and much
smaller than unity for all other states. Since the other states
decay much slower, at long times the Zeno state stops being the
dominant state in the decay process. Their weight in the
decomposition can be bounded by $\frac{d}{\gamma}$, and using the
Zeno state decay rate yields a crossover time of order
$\frac{\gamma}{u^2}\log(\frac{\gamma}{d})$ \cite{time_extension}.
The perturbation theory results are confirmed numerically, see Fig.
\ref{eigenvals}.

  \emph{B. Random matrix elements}.- When the disorder is sufficiently large,
  the matrix elements can be considered random \cite {brouwer1}. For simplicity,
  let us consider the case where the level spacing is constant, but the elements
  $u_\mu$ are randomly distributed around 0, with a standard
deviation $\bar{u}$.

 To understand the magnitudes of the matrix
elements involved in the physical realization of a small quantum dot
coupled to a larger disordered one which is coupled to a one channel
lead \cite{remark_about_single_channel_leads}, it is instructive to
look at their site representation.  If we denote the sites of the
larger dot by $|i \rangle$, then the isolated dot eigenstates are $|
\mu \rangle = \frac{1}{\sqrt{N}} \sum_ i \phi_i^\mu |i\rangle$,
where $\phi_i^\mu$ are random coefficients of order unity (assuming
the disorder is large enough, yet not too large as to make the
states localized). We shall assume that out of the $N$ sites, $S$
are coupled to $|0\rangle$, and $S'$ are coupled to the RC. Changing
basis to the set {$|\mu \rangle$}, one can verify that the couplings
to the RC and $|0 \rangle$ are random. If the matrix elements
$V_{j,k}$ (between a site $j$ and a state $k$ in the lead) do not
depend on $j$, after the lead elimination the Hamiltonian contains
terms $\sim |\mu\rangle \langle \nu| a_{\mu}a_{\nu}$, and by
multiplying the states $|\mu \rangle$ by a phase factor we can
obtain a model corresponding to Eq. (\ref{H}), with $\gamma_{\mu
\nu}$ real and positive. For simplicity we shall assume the
magnitude of $\gamma_{\mu \nu}$ to also be a constant, $\gamma$
\cite{remark_on_magnitude_fluctuations}. Denoting the typical matrix
element connecting a site in the larger dot with a site in the lead
by $v_0$ and the typical matrix element connecting a site in the
larger dot with a site in the single state dot $u_0$ , a
straightforward calculation shows that the typical tunneling matrix
elements in the Hamiltonian (\ref {H}) are given by the relations:
$\gamma = \gamma_0 \frac{\sqrt{S}}{N}$,
$\bar{u}\sim\frac{1}{\sqrt{N}}u_0$ where $\gamma_0= \nu_c |v_0|^2$
and $\bar{u}$ is the typical matrix element $u_\mu$ in Eq.
(\ref{H}). Notice that $u_0$, $v_0$,$\gamma_0$ are microscopic
parameters independent of the system size.

We shall now analyze the dynamics for four different cases: $u_0>
D$,  $u_0\sim D$, $D>u_0>d$, $d>u_0$.

\emph{Case I}.- To understand the behavior for $u_0> D$, we study
the degenerate case $D=0$, for which the Hamiltonian is $ H=
\sum_\mu u_\mu (|0 \rangle \langle \mu|+|\mu \rangle \langle 0|) -i
\gamma \sum_{\mu,\nu} |\mu \rangle \langle \nu| .$

Defining the states $|W_1\rangle = \frac{1}{\sqrt{\sum_\mu
u_\mu^2}}\sum_\mu u_\mu |\mu \rangle$ and $|W_3\rangle =
\frac{1}{\sqrt{N}} \sum_\mu |\mu \rangle $, the Hamiltonian takes
the form:

\be H=  {\sqrt{\sum_\mu u_\mu^2}}(|0 \rangle \langle W_1|+|W_1
\rangle \langle 0|) -iN\gamma |W_3 \rangle \langle W_3|.\ee

As $|W_1 \rangle$ and $|W_3\rangle$ are not orthogonal, it is useful
to use a Gram-Schmidt procedure to define $|W_2\rangle =
|W_1\rangle- |W_3\rangle \langle W_1 | W_3 \rangle $ and then
normalize it. Now we can represent the system's Hamiltonian in the
basis formed by $|0 \rangle$, $|W_2 \rangle$ and $|W_3 \rangle$ as a
3x3 matrix:

\be H=\left(%
\begin{array}{ccc}
  0 & U_2 \sqrt(1-c^2) & c U_2 \\
  U_2 \sqrt(1-c^2) & 0 & 0\\
  c U_2 & 0 & -i \gamma N  \\
\end{array}%
\right) ,\ee

where $U_1 \equiv {\sum_j u_j} \sim  u_0$, $U_2 \equiv \sqrt{\sum_j
u_j^2} \sim u_0 $, $c \equiv \frac{U_1}{U_2 \sqrt{N}} \sim
\frac{1}{\sqrt{N}}$.

Diagonalizing $H$ perturbativley in $\frac{1}{N}$ we find (using $c
\sim \frac{1}{\sqrt{N}}$) eigenvectors $|V_3\rangle=|W_3
\rangle+O(\frac{1}{N})$ with eigenvalue $-i \gamma_0
+O(\frac{1}{N})$, and $V_{\pm}=\frac{1}{\sqrt{2}} (|0\rangle \pm
|W_2 \rangle)$ with eigenvalues $\lambda_{\pm}= \pm U_2 -i \delta$,
where $\delta= -i\frac{U_1^2}{2 N^2 \gamma} \sim -i\frac{{u_0}^2}{2
N \gamma_0} .$

Since the projection of the initial state on $|V_3 \rangle$ is
negligible, the decay of the initial state is described by a
superposition of two exponentially decaying terms, with exponents $
\approx \pm {U_2} -i \delta$. This implies Rabi-type oscillations
(\cite {cohen_tanoudji_decay}, vol 1, p. 447) with a characteristic
frequency ${U_2} \sim u_0$, and with an envelope decaying
exponentially with rate $\frac{\bar{u}^2}{2 N \gamma} \sim
\frac{u_0^2}{2 N \gamma_0}$ \cite {tight_binding_model}. Notice that
in the case of finite bandwidth, the above analysis will be
approximately correct if $|\lambda_{\pm}| \gg D$ and $\gamma \gg d$
. The first condition gives the restriction $D \ll u_0$. Since this
relation is $N$ independent, both cases are physically accessible in
the limit of large $N$. The result of a numerical simulation is
shown in Fig. \ref {random3}(a).

 Notice two peculiarities of the result: First, as in Eq.~(\ref {generalized_eq}),
when $\gamma$ is increased the decay is slower. Second, the decay is
much slower than in the ordered case, by a factor of $2N$. This can
be understood as follows: the sum of all imaginary parts of the
eigenvalues exactly equals $-iN \gamma$. Since the Dicke eigenvector
is still of the same form as before, with an eigenvalue
approximately given by $-i N \gamma$, the rest of the eigenvalues
have small imaginary parts. Plugging the Dicke eigenvalue into Eq.
(\ref {eigenval_eq}) as $-i N \gamma + \epsilon$, one finds that
$\epsilon \approx \frac{i (\sum u_j)^2}{N^2 \gamma}$. For constant
matrix elements, the part of the decay sum rule not taken up by the
Dicke eigenvector, namely $i \frac{v^2}{\gamma}$, was mostly taken
up by the Zeno eigenvector. Here, the sum of all other decay rates
$\sim \frac{\bar{u}^2}{\gamma N}$, distributed among \emph{all}
eigenvalues.

\emph{Case II}.- For $D \sim u_0$ the 'disorder' is not
'self-averaging', and the decay depends on the realization of the
disorder. An example is shown in Fig. \ref {random3}(b). The initial
smooth decay is approximately a sinc (which, in the general case,
will be replaced by the Fourier transform of the density of states),
as shown in the inset, up to times of order $\frac{1}{D}$, in which
the amplitude decays almost to 0. Then random oscillations come into
play, with an amplitude of order $\frac{1}{\sqrt {N}}$.

One can verify that the real parts of the eigenvalues form a band of
bandwidth $D$ with constant density, by considering the eigenvalue
Eq. (\ref {eigenval_eq}). If $\Sigma_1$ happens to be exactly
imaginary, then it is easy to see graphically that there is a
solution $\lambda$ between each consecutive energies~$E_\mu$. Using
Eq. (\ref {eigenvectors}), one can now check self-consistently, that
$\Sigma_1$ is almost purely imaginary, since $\gamma \sum
\frac{1}{\lambda - E_\mu} \gg 1$. Therefore the eigenvalues are
almost real, and are distributed with constant as claimed above.
Since the eigenvectors are nearly real, so are the coefficients in
Eq. (\ref{time_evolution}), but these are now randomly distributed.
We have the interference of $N$ oscillating components with
positive, random coefficients. Then the amplitude is described by $
A_0(t) \approx \sum_n q_n^2 e^ {-i \omega_n t} ,$ where $q_n$ are
independent, random variables. To understand the behavior in this
case, it is instructive to look at the ensemble average of
probability, denoted by an overline. 
This gives:
$$ \overline{|A_0(t)^2|}> \approx \sum_{n,m} <q_{n}^2 q_m^2> e^ {-i \omega_n
t} e^ {i \omega_{m} t} . \label {sinc}$$

Performing the sum under the assumption $t d \ll 1$ gives $
|A_0(t)|^2 \sim \rm{sinc}(Dt)$, which is demonstrated in Fig. \ref
{random3}(b). Although we have random oscillations, a characteristic
frequency seems apparent. This is because typically a few states
are, by chance, coupled more than the others. Since in our case $q_n
\sim |\langle V_n |0 \rangle| \sim \frac{1}{\sqrt{N}}$ the amplitude
of the oscillations $\sim \frac{1}{\sqrt{N}}$ as well.

\emph{Case III}.- Defining $\lambda_0=\pi {\bar {u}^2}/{d} \sim
{u_0^2}/{D}$, the FGR rate, we find different behavior for the cases
$D \gtrless \lambda_0$. Notice that the condition of the crossover
$D \sim \lambda_0$ is the same as before, namely $D \sim u_0$.

If $D \gg \lambda_0$ ($D \gg u_0$), a coefficient of an eigenvalue
$\lambda$ will have a typical size
$\frac{1}{1+(\frac{\lambda}{\lambda_0})^2}$. Notice that this fits
the exact sum $\sum C_n^2=1$.
 Thus, we have a superposition of $M \sim \pi \frac{\bar
{u}^2}{d^2} $ oscillatory signals, with (positive) random
coefficients. If furthermore $\bar{u} \gg d$ (which is equivalent to
$u_0 \sqrt{N} \gg D$), we will have a large number of random
components within a complete Lorentzian, and therefore the FGR
exponential decay will be retrieved (but with the Zeno effect
suppressed).

\emph{Case IV}.- For $u_0 \ll d$, a single state is relevant, and
the system will show (decaying) Rabi oscillations.

\emph{Conclusions}.- We considered the generic problem of a single
state coupled to a real continuum via a pseudo continuum. In the
ordered case, we found that a single eigenvector characterizes most
of the decay of the system, and the decay becomes slower when
increasing the coupling to the real continuum. When the bandwidth
$D$ is smaller than the typical matrix element $u_0$, adding
disorder causes the decay to be much slower, and introduces
oscillations in time. When $D \sim u_0$, mesoscopic fluctuations in
the probability to stay in the single state as a function of time
follow, while for a larger bandwidth, FGR exponential decay is
retrieved.

We thank R. J. Schoelkopf and S. M. Girvin for pointing out
\cite{purcell}, P. W. Brouwer and P. S. Silvestrov for useful
discussions, and F. von Oppen and S. A. Gurvitz for their important
remarks. This work was supported by a BMBF DIP grant as well as by
ISF grants and the Center of Excellence Program.

\bibliographystyle {prsty}

\begin{thebibliography}{10}
\bibitem{remark_about_single_channel_leads}
For a single channel lead (as defined by \cite{zelevinksy_unstable})
the matrix
  element between a state in the PC and in the RC does not depend on the index
  of the RC state, provided the coupling region is smaller than the typical
  wavelength \cite {raikh}. The case of multi channel leads will be treated in
  future works.
\bibitem{remark_about_alternating_sign_case}
When the matrix elements have alternating sign, the Zeno effect will
not appear
  \cite {gurvitz1}, and FGR holds.
\bibitem{quantum_dot_physical_realization}
Consider a number of small quantum dots with no tunneling amongst
them, at
  equal distances from the single quantum dot. The matrix elements $u_j$ will
  be constant.
\bibitem{remark_about_non_hermitian}
To find the coefficients $C_n$, it is helpful to note that since $H$
is
  symmetric, the scalar product of two different eigenvectors (without taking
  the complex conjugate of one of them) vanishes. It follows that if we
  redefine the projection operation as a scalar product, and normalize the
  eigenvectors accordingly, the usual formulas for the eigenvector
  decomposition will hold.
\bibitem{time_extension}
Notice that the upper time for which the decay is exponential
increases with
  $\gamma$. For, roughly, $\gamma > \gamma_0$, where $\gamma_0$ is the FGR
  rate, the coupling to the continuum increases the exponential decay regime.
\bibitem{remark_on_magnitude_fluctuations}
We have confirmed that fluctuations in the magnitude do not lead to
a
  qualitative difference in the results in most cases, as long as the sign of
  the matrix elements is constant.
\bibitem{tight_binding_model}
In the tight-binding picture with $S=1$ the rate has a simple
interpretation,
  being the typical matrix element between the SS and the site it is coupled to
  in the larger dot.
\end{thebibliography}

\end{document}